\documentclass[12pt]{article}
\usepackage{fullpage}

\usepackage{amsmath}
\usepackage[psamsfonts]{amssymb}
\usepackage{amsthm}
\usepackage{xspace}

\newcommand{\Span}{\mathrm{Span}}

\newcommand{\C}{{\mathbb C}}
\newcommand{\Z}{{\mathbb Z}}

\newcommand{\N}{{\mathbb N}}

\newcommand{\uu}{{\mathbf u}}
\newcommand{\vv}{{\mathbf v}}
\newcommand{\jj}{{\mathbf j}}

\newcommand{\zz}{{\mathbf z}}
\newcommand{\hh}{{\mathbf h}}
\newcommand{\PP}{{\mathbb P}}

\newcommand{\F}{{\mathcal H}}
\newcommand{\OO}{{\mathcal O}}
\newcommand{\V}{{\mathcal V}}

\newcommand{\abs}[1]{\lvert#1\rvert}

\newcommand{\IP}[1]{\langle#1\rangle}
\newcommand{\iso}{\simeq}
\newcommand{\done}{$\hfill \hfill \blacksquare$ \bigskip}
\newcommand{\dwrt}[1]{\frac{\partial}{\partial#1}}

\newcommand{\ci}{\oint}
\newcommand{\Div}{\mathrm{div}\ }
\newcommand{\Mod}{\mathrm{mod}\ }

\newcommand{\chit}{\tilde{\chi}}
\newcommand{\etat}{\tilde{\eta}}

\newtheorem{defn}{Definition}
\newtheorem{thm}[defn]{Theorem}
\newtheorem{lem}[defn]{Lemma}

\newcommand{\oproof}[1]{\noindent {\bf Proof#1.\ }}

\newcommand{\ti}[1]{\textit{#1}}

\begin{document}

\begin{flushright}
  DAMTP-2000-50
\end{flushright}

\begin{center} \LARGE{\textbf{Zhu's Theorem and an algebraic characterization of chiral blocks}} \end{center}
\vskip 1.0cm
\begin{center}{\large Andrew Neitzke\footnote{aneitzke@alumni.princeton.edu}\\
DAMTP, Cambridge University\\
Cambridge CB3 0WA\\
United Kingdom\\}\end{center}

\begin{center} May 16, 2000 \end{center}

\vskip 0.8cm

\begin{abstract}

Working in the axiomatic framework recently proposed by Gaberdiel and Goddard, we prove a generalized version
of Zhu's Theorem; for any chiral bosonic conformal field theory on the sphere, our result characterizes 
the chiral blocks in terms of a certain quotient of the Fock space.  We also establish,
under a finiteness hypothesis closely related to rationality of the theory, that the
relevant Knizhnik-Zamolodchikov-type equation admits solutions.

\end{abstract}

\section{Introduction} \label{intro}

The problem of determining the chiral blocks in a given 
conformal field theory is \ti{a priori} a difficult one.  In certain 
specific cases this problem has been
completely solved -- e.g., for a broad class of well behaved theories, the chiral blocks
are understood to arise from ``Feynman diagrams'' with only $3$-valent vertices, with the
interaction vertices completely determined by the
fusion rules \cite{DFMS}.  Furthermore the spaces of chiral blocks have been computed explicitly
in certain cases, e.g. the WZW models, for which they have a straightforward
algebraic description as spaces of coinvariants \cite{FSV}, \cite{KL}.  

For more general theories (even rational ones), somewhat less is known.
The most significant progress was made by Zhu,
who in \cite{Zhu} introduced a completely algebraic technique for determining the highest weight representations of
a vertex operator algebra; namely, he constructed a functor which associates to a vertex operator
algebra $V$ (with conformal weights in $\N$) 
an associative algebra $A(V)$,
such that the irreducible representations of $A(V)$ are in 1-1 correspondence
with the irreducible highest weight representations of $V$.  In fact, given a representation of $A(V)$,
Zhu constructed the corresponding representation
of $V$ by first defining correlation functions on the sphere and then factorizing to obtain the states;
so when $V$ is the algebra of fields in the vacuum sector of a chiral conformal field theory, we can interpret
Zhu's construction as giving all the $2$-point chiral blocks.
Roughly speaking, $A(V)$ is the algebra of zero
modes of $V$ (see \cite{NB} for a discussion of this point, \cite{DLM}, \cite{Li} for other
facets of $A(V)$, and \cite{FZ}, \cite{Lucke}, \cite{DLM2} for some explicit calculations).
A modification of Zhu's construction allows one
to compute, in a similar algebraic fashion, the fusion rules of the
theory \cite{FZ}.

This paper is mainly concerned with a generalization of Zhu's construction
to the case of $k$-point conformal blocks.  
We now turn to a description of its contents.

In Section \ref{notation} we fix notation
and briefly review the formalism introduced by Gaberdiel and Goddard in
\cite{GG}, which is a convenient framework in which to state the theorems
of this paper.

In Section \ref{zhusec} we discuss Zhu's construction and a natural generalization, 
first mentioned in \cite{GG},
which associates to $V$ and any $\uu = (u_1, \dots, u_k) \in \PP^k$ a vector space $A_\uu$.  This $A_\uu$ will be
obtained as a quotient of the Fock space at $0 \in \PP$.

In Section \ref{zhugen} we prove that, when the $u_i$ are distinct, any linear functional $\eta \in (A_\uu)^*$ 
corresponds to the value of
a chiral block at $\uu$, in the sense that $\eta$ induces a consistent
prescription for correlation functions
$\IP{\prod_{i=1}^k \phi_i(u_i) \cdots}$
where the dots indicate arbitrary insertions of the vertex operators in $V$.
So $\eta$ corresponds to a particular way of coupling some set of
$V$-primary fields $\phi_i$ placed at the points $u_i$.

In Section \ref{funcdep} we introduce a certain finiteness condition on $V$ which generalizes
Zhu's ``condition $C$.''  Under this condition, which appears to be closely related to rationality of $V$, we show that
the Knizhnik-Zamolodchikov-type equation governing the $\uu$-dependence of the chiral block (obtained
essentially by making the substitution $L_{-1} \to \partial$) admits
a solution.

Finally, in Section \ref{discussion} we discuss remaining open questions,
and a possible relation between
the present work and the Friedan-Shenker vector bundle formalism
\cite{FS}.

\section{Hypotheses and notation} \label{notation}

We assume the reader is familiar with basic notions of conformal field
theory, as found for instance in \cite{DFMS}, \cite{Greview}.  Some
acquaintance with the language of vertex operator algebras \cite{FLM},
\cite{Kac}, \cite{Borch} is also helpful.

At all times in this paper we are considering a fixed chiral bosonic conformal field theory
on the sphere $\PP$.  To be completely rigorous, by a ``chiral bosonic conformal field theory'' we mean
an object of the type discussed in \cite{GG}.  The details of the chosen formalism are generally not
essential to following the ideas of this paper, however; what is essential is just that a chiral conformal field
theory on $\PP$ is regarded as defined by its amplitudes.
We write these amplitudes $\IP{\prod_{a=1}^k V(\psi_a, z_a)}$, with the vertex operator corresponding
to $\psi$ written $V(\psi, z)$
and its modes written 
\begin{equation}
V(\psi, z) = \sum_{n \in \Z} V_n(\psi) z^{-n-h_\psi}
\end{equation}
(so the grading is by conformal weight, which we always represent by the letter $h$).
The only exception to this rule is the Virasoro field which we write $L(z)$, with modes $L_n$.
The space of Virasoro quasiprimary states is denoted by $V$; it has a grading by conformal weight, which we assume
can be taken of the form
$V = \oplus_{h=0}^\infty V^h$, with each $V^h$ finite-dimensional.

In \cite{GG} the role of ``space of states'' 
is played by a collection of topological vector spaces denoted $\V^\OO$,
for $\OO$ an open set in $\PP$ (usually with $\PP - \OO$ simply connected.)  These vector spaces are obtained by
factorization from the amplitudes, with two states regarded as equal 
just if they agree in correlation functions with vertex operators
inside $\OO$.
Roughly speaking, an element $\chi \in \V^\OO$ is a coherent state (or limit of coherent states)
constructed out of fields away from $\OO$.  In this paper we
make explicit 
reference to $\V^\OO$ only occasionally; when we do, $\OO$ will always have simply connected complement.
We will make frequent use of the Fock space $\F$ at $0$,
which is defined similarly by factorization.

We will often consider meromorphic functions and differentials defined on the Riemann sphere $\PP$. 
It is convenient to use the language of ``divisors'' (see \cite{GH}) to classify the zeros
and poles of these functions.
So let a divisor on $\PP$ be any formal sum of the form
\begin{equation}
D = \sum_{P \in \PP} c_P [P], \qquad c_P \in \Z, \qquad \textrm{finitely many } c_P \ne 0.
\end{equation}
Divisors can be added in the obvious way.  We say $D \ge 0$ if all $c_P \ge 0$.  
Now let $\nu_P(f)$ denote the order of vanishing of $f$ at $P$, and define the divisor of $f$ to be
\begin{equation}
\Div f = \sum_{P \in \PP} \nu_P(f) [P],
\end{equation}
so that $\Div f \ge -D$ if the poles of $f$ are ``at worst given by $D$.''  Clearly $\Div fg = \Div f + \Div g$.
We can similarly define $\Div \omega$ where $\omega$ is a meromorphic $k$-differential on $\PP$; explicitly,
such an $\omega$ can always be written as $\omega = f dz^{\otimes k}$ for some $f$, and then
we have 
\begin{equation}
\Div \omega = \Div f + k \Div dz = \Div f - 2k [\infty].
\end{equation}
The crucial analytic property which the amplitudes of the theory must possess is that
for $\IP{V(\psi, z) \cdots} dz^{\otimes h_\psi}$ is meromorphic on $\PP$ for any $\psi \in V$, and has poles only when $z$ meets
the coordinates of the insertions $\cdots$.

\section{Zhu's subspace} \label{zhusec}

Zhu in \cite{Zhu}
introduced a purely algebraic mechanism for determining the highest weight 
representations of a chiral theory.  This construction, when generalized
to $k$-point functions, amounts to the following:  Fix $\uu \in (\PP - \{0\})^k$ and 
consider a subspace $O_\uu \subset \F$ of the Fock space, defined by
\begin{equation} \label{defO}
\begin{split}
O_\uu = \Span \ \Big\{ \ci_0 dz g(z) V(\psi, z) \chi \ \Big\vert \ &\chi \in \F, \psi \in V, \\
                    & \Div g dz^{\otimes -h_\psi+1} \ge -N[0] + \sum_{i=1}^k h_\psi [u_i] \textrm{    for some } N\Big\}.
\end{split}
\end{equation}
This definition can be motivated in the following way:  
as described in \cite{GG}, define a ``highest weight state at $\uu$'' to be 
any state $\Sigma$ such that
\begin{equation} \label{highwtcond}
\Div \left( \IP{ \Sigma V(\psi, z) } dz^{\otimes h_\psi} \right) \ge \sum_{i=1}^k -h_\psi [u_i], \qquad \forall \psi \in \F.
\end{equation}
(Informally, the idea is that $\Sigma$ should stand for insertions
of primary fields $\phi_1(u_1) \cdots \phi_k(u_k)$; then the requirement
\eqref{highwtcond} just says that each $\phi_i$ is annihilated by positive
modes of $V(\psi, z)$.) 
If we fix such a $\Sigma$ and consider $\omega \in O_\uu$, then $\IP{\Sigma \omega}$ must vanish, 
because substituting
the integral appearing in 
\eqref{defO} for $\omega$ we find that the resulting integrand has been engineered to have no poles
on $\PP - \{0\}$.  $O_\uu$ therefore represents a space of 
``states in the Fock space at $0$ 
which are orthogonal to primary fields placed at the points $u_1, \dots, u_k$.''

For most of this paper we consider only the case where all $u_i$ are distinct (though the behavior
of $O_\uu$ when some of the $u_i$ come together is important --- in particular, it motivates
the finiteness condition we impose in Section \ref{funcdep}, and also see Section \ref{discussion}.)

Now suppose given a particular $\Sigma$ which is a highest weight state at $\uu$.  Then $\Sigma$ induces
a linear functional $\eta: \F \to \C$ by the rule
$\eta(\chi) = \IP{\Sigma \chi}$, and we have just seen that this $\eta$ must vanish
on $O_\uu$.  Conversely, in the special case $k=2$, Zhu essentially showed that 
any $\eta: \F \to \C$ vanishing on $O_\uu$
in fact comes from a $\Sigma \in \V^\OO$ satisfying \eqref{highwtcond} --- in other words, every such $\eta$ comes from a 
highest weight representation.
So we have a correspondence between linear functionals on $\F / O_\uu$ and representations of our theory.  Actually, different linear functionals
can give rise to equivalent representations; the precise formulation in \cite{Zhu}
defines an algebra structure on the quotient $\F / O_\uu$ and shows that the irreducible 
representations of this algebra are
exactly the highest weight representations of the theory.  This is a 
remarkable result as it provides a completely systematic way of constructing
the representations, which are \ti{a priori} rather complicated objects from an algebraic standpoint 
since one needs to specify the action of every field in the theory.

In calculations it will be useful to know that the function $g$ in \eqref{defO} can be chosen to
have an ancillary property, namely, we can choose it to be holomorphic in $\uu$:  e.g. it is easily checked
that
\begin{equation} \label{fchoice}
g_N(z) = z^{-(N+1+(k-2)h)} \prod_{i=1}^k (z-u_i)^h
\end{equation}
satisfies the condition in \eqref{defO} for all $N \ge 1$.  Furthermore, a straightforward induction shows that it is
actually sufficient to use only the $g_N$ in the definition of $O_\uu$ 
(essentially because a function satisfying \eqref{defO} is determined
by its singular part at $0$.)

\section{A generalization of Zhu's Theorem} \label{zhugen}

We will now give a generalization of Zhu's result mentioned above ---
which can be viewed as a construction of correlation functions
corresponding to insertions of representations of $A(V)$ at $2$ 
points --- to general $k$-point functions.

To prove our generalized version of Zhu's Theorem we need to construct correlation functions 
which induce a given linear $\eta: \F \to \C$ (vanishing on $O_\uu$).
We represent these (putative) correlation functions by the notation
\begin{equation} \label{defcf}
\IP{\prod_{a=1}^l V(\psi_a, z_a)}_\eta,
\end{equation}
where the subscript $\eta$ reminds us that these are not the
vacuum correlation functions.
In a sense we have no choice in defining the functions \eqref{defcf},
because to say they are induced from $\eta$ is exactly to say that
$\eta$ already defines for us their Laurent series; the real question is whether these series converge.  However, working with these
Laurent series will be somewhat difficult, because the correlation functions are 
expected to have poles whenever two
of the $z_i$ coincide, and the series in question are expanded about the point $\zz = (0, \dots, 0)$.
We must therefore be careful about the domain in which we are working.
We will use the letter $R$ to denote one of the $l!$ 
possible permutations of the coordinates $z_1, \dots, z_l$; by
abuse of notation, $R$ is the region $\{\zz: \abs{z_{R(1)}} > \abs{z_{R(2)}} > \cdots > \abs{z_{R(l)}}\}$.

So given $\eta$, some $(\psi_1, \dots, \psi_l) \in V^{\otimes l}$, and a region $R$, we define a formal power series by
\begin{equation} \label{defseries}
\IP{\prod_{a=1}^l V(\psi_a, z_a)}_{\eta,R} = \sum_{\jj \in \Z^l} \eta \left( \prod_{i=1}^l V_{j_{R(i)}}(\psi_{R(i)}) \right) \zz^{- \jj - \hh}
\end{equation}
where $\jj$ is a multi-index, so e.g. by $\zz^{-\jj-\hh}$ we mean $\prod_{i=1}^l z_i^{-j_i-h_i}$.
For $\eta$ which could be induced from primary fields inserted at $\uu$ --- in other words, $\eta$ vanishing on the
subspace $O_\uu$ introduced in Section \ref{zhusec} --- we will show that the power series \eqref{defseries} 
are the Laurent expansions of a single function 
$\IP{\prod_{a=1}^l V(\psi_a, z_a)}_\eta$ in the different regions $R$.  This is the content of Theorem
\ref{bigThm}, toward which we are working (the next three lemmas are somewhat technical, so the reader may
want to flip to the theorem first.)

In order to prove Theorem \ref{bigThm} we first
establish that the series \eqref{defseries} obey a formal version of the operator product expansion.  To
formulate this statement precisely we need one more bit of notation:  for a meromorphic function $f(\zz)$
with poles only at $z_i = z_j$, let $\iota_{i,j} f$ mean ``the Laurent
series for $f$ around $(0, \dots, 0)$, expanded in the region where $\abs{z_i} > \abs{z_j}$'' (this
notation is commonly employed in the study of vertex operator algebras, see e.g. \cite{Kac}, \cite{FLM}).
Then we can state
\begin{lem} \label{opelem} Let $R = \{\abs{z_1} > \dots > \abs{z_l}\}$ (for simplicity).
Then for any $\eta: \F \to \C$ the power series defined by $\eta$ obey the following ``operator
product expansion'' identities:
\begin{itemize}
\item For all $m\in[1,l)$, we have the OPE as $z_m \to z_l$:
\begin{equation} \label{ope}
\begin{split}
\IP{\prod_{a=1}^l & V(\psi_a, z_a)}_{\eta,R} =  \\
 & \sum_{i=m+1}^{l-1} \sum_{n = - h_1 - h_i}^{-1} (\iota_{m,i} - \iota_{i,m}) (z_m - z_i)^{n} \IP{V(V_{-n-h_m}(\psi_m) \psi_i, z_i) \prod_{a \ne m,i} V(\psi_a, z_a)}_{\eta,R} \\
+ &\sum_{n = -h_m - h_l}^\infty \iota_{m,l} (z_m - z_l)^n \IP{V(V_{-n-h_m}(\psi_m) \psi_l, z_l) \prod_{a \ne m,l} V(\psi_a, z_a)}_{\eta,R}.
\end{split}
\end{equation}
\item We also have an OPE as $z_l \to z_{l-1}$:
\begin{equation} \label{opeshort}
\begin{split}
\IP{\prod_{a=1}^l & V(\psi_a, z_a)}_{\eta,R} = \\
&\sum_{n = -h_{l-1} - h_l}^\infty \iota_{l,{l-1}} (z_l - z_{l-1})^n \IP{V(V_{-n-h_l}(\psi_l) \psi_{l-1}, z_{l-1}) \prod_{a \ne {l-1},l} V(\psi_a, z_a)}_{\eta,R}.
\end{split}
\end{equation}
\end{itemize}
\end{lem}

\oproof{} The essence of the proof is the observation that checking any coefficient in the above identities involves
only a finite computation; this computation amounts to verifying that a 
certain state $\chi \in \F$ is annihilated by $\eta$.  But the
OPE of the conformal field theory then shows that 
this same $\chi$ is annihilated by the linear functionals induced by correlations with vertex operators inserted
away from $0$.  Then by the factorization property $\chi = 0$, so naturally $\eta(\chi) = 0$.
In other words:
the operator product expansion is already encoded in 
the definition of $\F$, so naturally every linear functional on $\F$ must obey it.

More explicitly:  first we prove \eqref{ope}.  For notational simplicity we consider only the case $m=1$.
Fix a multi-index $\jj$ and consider the coefficient of $\zz^\jj$ in \eqref{ope}; we have to show this
coefficient receives only finitely many contributions on each side.  The left side manifestly has only a single
term involving $\zz^\jj$.  On the other hand, each term in the sums on the right side can contain at
most one contribution to the coefficient of $\zz^\jj$.  The double sum contains only finitely many terms, so manifestly
makes only a finite contribution.  The single sum contains infinitely many terms; indeed,
replacing $(z_1 - z_l)^n$ and $\IP{V(V_{-n-h_1}(\psi_1) \psi_l, z_l) \prod_{a \ne 1,l} V(\psi_a, z_a)}_{\eta,R}$ by their power series expansions,
we find that for each $k \ge 0$ the coefficient of $\zz^\jj$ can receive a contribution proportional to
\begin{equation} \label{propterm}
\eta \left( \left( \prod_{i=2}^{l-1} V_{-j_i-h_i}(\psi_i) \right) V_{- j_1 - j_l - h_1 - h_l}( V_{j_1 - h_1 - k}(\psi_1) \psi_l) \Omega \right).
\end{equation}
So we need to show that only finitely many of these terms are nonzero.
When $\jj$ is fixed, then setting $\alpha = j_1 + j_l + h_1 + h_l, \beta = h_1 + h_l - j_1,$ we find that 
\eqref{propterm} depends linearly on a state of the form $V_{-\alpha}(\chi_k) \Omega$ where
$\chi_k$ has weight $\beta + k$.  But such an expression always vanishes when $k > \alpha-\beta$.
(Note that we are taking advantage of a special property of the vacuum, and so 
this only works because $z_l$ is the coordinate closest to the origin --- this is the 
reason why we restricted ourselves to that special case.)  
Hence the coefficient of $\zz^\jj$ receives only finitely
many contributions on each side of \eqref{ope}, and we can rewrite \eqref{ope} in the form
\begin{equation} \label{ope2}
\eta(\chi_\jj) = 0
\end{equation}
for some $\chi_\jj \in \F$.

Now choose some $\OO \subset \PP$ containing $0$.  Then the space $\V^\OO$ of limits of coherent states 
is embedded in $\F^*$ as a dense 
subspace (in the weak-$*$ topology induced from $\F$ --- see \cite{GG}).
But for $\eta$ induced from $\Sigma \in \V^\OO$, the power
series we are considering actually do converge --- $\IP{\prod_{a=1}^l V(\psi_a, z_a)}_{\eta, R}$ is exactly the power series expansion
of $f(\zz) = \IP{\Sigma \prod_{a=1}^l V(\psi_a, z_a)}$ for $\zz \in R$.  To determine this expansion we make use of
a construction from \cite{Gaberdiel}, as follows.  Fix $z_2, \dots, z_l \in \OO$ with $\abs{z_2} > \cdots > \abs{z_l}$ and 
note that for $z_1$ close enough to $z_l$ we have \cite{DFMS}, \cite{Greview}
\begin{equation}
f(\zz) = \sum_{n=-h_1-h_l}^\infty (z_1 - z_l)^{n} \IP{\Sigma V(V_{-n-h_1}(\psi_1) \psi_l, z_l) \prod_{a \ne 1,l} V(\psi_a, z_a)}.
\end{equation}
From the operator product expansion we also know the pole structure of the correlation function, so that if we put
\begin{equation}
g(\zz) = \sum_{i=2}^{l-1} \sum_{n = - h_1 - h_i}^{-1} (z_1 - z_i)^{n} \IP{\Sigma V(V_{-n-h_1}(\psi_1) \psi_i, z_i) \prod_{a \ne 1,i} V(\psi_a, z_a)}
\end{equation}
then $f(\zz) - g(\zz)$ has no poles as a function of $z_1 \in \OO$.  Its power series expansion about $z_1 = z_l$ therefore converges on any disc contained in $\OO$.
So we can write $f(\zz)$ when $\abs{z_1} > \abs{z_2}$ as (the expansion of $f(\zz) - g(\zz)$ for $z_1$ near $z_l$) plus 
(the expansion of $g(\zz)$ for $\abs{z_1} > \abs{z_2}$).  This gives exactly \eqref{ope} except that we have substituted fixed values for $z_2, \dots, z_l$;
but since those values were arbitrary we get $\eqref{ope}$ as an identity of functions, which implies the desired identity of power series.
So \eqref{ope}, and hence \eqref{ope2}, hold for all $\eta \in \V^\OO$.  Then \eqref{ope2} and \eqref{ope} must
hold for all $\eta \in \F^*$, completing the proof of \eqref{ope} (in the case $m=1$, but the other $m$ are proven in an exactly analogous way.)

The proof of \eqref{opeshort} is similar to the proof of \eqref{ope} in case $m=l-1$ (which is actually somewhat easier than the general case
because there are no poles to be subtracted.)  The point is that the necessary finiteness condition will hold on the right side of \eqref{opeshort},
because after we decompose the field at $z_l$ into modes acting at $z_{l-1}$, the field at $z_{l-1}$ will be the one closest to the origin; so
we can argue as above. \done

Fix $k$ and fix some $\uu = (u_1, \dots, u_k) \in (\C-\{0\})^k$, with all $u_i$ distinct.  Our strategy in proving
Theorem \ref{bigThm} will be to first establish convergence of modified power series
in which we have shifted the poles from $z_i = u_j$ to $z_i = \infty$; once this is established the rest
is easy.  The next two lemmas concern these modified power series.

\begin{lem} \label{pL} Fix $l \ge 0$ and let $R = \{\abs{z_1} > \dots > \abs{z_l}\}$.  Suppose given $\eta: \F \to \C$ such that $O_\uu \subset \ker \eta$, and
$(\psi_1, \dots, \psi_l) \in V^{\otimes l}$.  Then in the power series
\begin{equation} \label{powr}
\left( \prod_{m=1}^k (z_1 - u_m)^{h_1} \right) \IP{\prod_{a=1}^l V(\psi_a, z_a)}_{\eta,R}
\end{equation}
the coefficient of $\zz^\jj$ vanishes whenever $j_1 > (k-2) h_1$.
\end{lem}
\oproof{} The motivating idea is that as a function of $z_1$ in the region $R$,
$\IP{\prod_{a=1}^l V(\psi_a, z_a)}_\eta$ should only have poles at $z_1 = u_i$, of order at 
most $h_1$.  By multiplying by $\prod_{m=1}^k (z_1 - u_m)^{h_1}$ we convert all these poles to a single pole at $\infty$, still
of bounded order; and since all other poles have been removed, the power series expansion in $R$ of the
resulting function can be expected
to converge up to $z_1 = \infty$.  The bounded order of the pole at $\infty$ will therefore be manifested as a cutoff in
the power series.

Explicitly, the proof consists in noting that the coefficient of $\zz^\jj$ in \eqref{powr} is given by
\begin{equation}
\eta \left( \ci_0 \frac{dz_1}{z_1^{j_1+1}} \prod_{m=1}^k (z_1 - u_m)^{h_1} V(\psi_1, z_1) \prod_{a=2}^l V_{-j_a - h_a}(\psi_a) \Omega \right)
\end{equation}
which vanishes by hypothesis for $j_1 > (k-2) h_1$. \done

Now we are in a position to show that our modified power series actually converge.
\begin{lem} \label{modconv} Suppose given $\eta: \F \to \C$ such that $O_\uu \subset \ker \eta$.  Fix $l \ge 0$, and for
any $(c_1, \dots, c_l) \in (\Z^+)^l$, define
\begin{equation}
\Pi = \prod_{j=1}^l \prod_{m=1}^k (z_j - u_m)^{c_j}.
\end{equation}
Then for any $(\psi_1, \dots, \psi_l) \in V^{\otimes l}$, the $c_j$ can be chosen sufficiently large that the power series
\begin{equation}
\Pi \cdot \IP{\prod_{a=1}^l V(\psi_a, z_a)}_{\eta,R}
\end{equation}
is convergent in $R$.  Furthermore, this power series can be continued to a meromorphic function defined on $\C^l$, with poles
only at $z_i = z_j$, independent of $R$.  For $l>1$ this function is given recursively by the formula
\begin{equation} \label{ope5}
\begin{split}
\Pi \cdot \IP{\prod_{a=1}^l V(\psi_a, z_a)}_{\eta}
= & \Pi \cdot \sum_{i=2}^{l} \sum_{n = - h_1 - h_i}^{-1} (z_1 - z_i)^{n} \IP{V(V_{-n-h_1}(\psi_1) \psi_i, z_i) \prod_{a \ne 1,i} V(\psi_a, z_a)}_{\eta} \\
- & \Pi \cdot \sum_{i=2}^{l-1} \sum_{n = - h_1 - h_i}^{-1} \iota_{i,1} (z_1 - z_i)^{n} \IP{V(V_{-n-h_1}(\psi_1) \psi_i, z_i) \prod_{a \ne 1,i} V(\psi_a, z_a)}_{\eta} \\
+ & \Pi \cdot \sum_{n=0}^\infty (z_1 - z_l)^n \IP{V(V_{-n-h_1}(\psi_1) \psi_l, z_l) \prod_{a \ne 1,l} V(\psi_a, z_a)}_{\eta},
\end{split}
\end{equation}
which must be interpreted as follows:  the first term is a meromorphic function in $z_1$,
and the second two together define a convergent power series in $z_1$ (in fact a polynomial.)
\end{lem}

\oproof{} By induction on $l$.  For $l=1$, choose $c_1=h_1$; then we are just considering
\begin{equation}
\prod_{i=1}^k (z - u_i)^{h} \IP{V(\psi, z)}_{\eta,R}
\end{equation}
and Lemma \ref{pL} says this series can have no power of $z$ exceeding $z^{(k-2)h}$.  On the other hand,
from the definition we see immediately that it can have no pole at $z=0$; so in this case the series is just a polynomial. 

So take $l \ge 2$ and assume the lemma true for $l-1$.  We use the fact that the power series satisfy a formal OPE as
$z_1 \to z_l$ (Lemma \ref{opelem}) to reduce a correlator with $l$ vertex operators to a sum of
correlators with $l-1$ vertex operators.  From the case $m=1$ of \eqref{ope}, we have
the expansion
\begin{equation} \label{ope3}
\begin{split}
\IP{\prod_{a=1}^l V(\psi_a, z_a)}_{\eta,R} = \sum_{i=2}^{l-1} &\sum_{n = - h_1 - h_i}^{-1} (\iota_{1,i} - \iota_{i,1}) (z_1 - z_i)^{n} \IP{V(V_{-n-h_1}(\psi_1) \psi_i, z_i) \prod_{a \ne 1,i} V(\psi_a, z_a)}_{\eta,R} \\
+ &\sum_{n = -h_1 - h_l}^\infty \iota_{1,l} (z_1 - z_l)^n \IP{V(V_{-n-h_1}(\psi_1) \psi_l, z_l) \prod_{a \ne 1,l} V(\psi_a, z_a)}_{\eta,R},
\end{split}
\end{equation}
where $R = \{\abs{z_1} > \dots > \abs{z_{l}}\}.$  To make the notation more palatable we
now define, for $i \in (1,l]$ and $n \in \Z$,
\begin{equation}
g_{i,n}(z_2, \dots, z_l) = \left( \prod_{j=2}^l \prod_{m=1}^k (z_j - u_m)^{c_j} \right) \IP{V(V_{-n-h_1}(\psi_1) \psi_i, z_i) \prod_{a \ne 1,i} V(\psi_a, z_a)}_{\eta,R}.
\end{equation}
In addition, we fix the $c_j$ ($j \in (1,l]$) sufficiently large that $g_{i,n}$
is convergent in $R$ for $(i \in (1,l)$, $n \in [-h_1-h_i, -1])$ and for $(i=l, n \in [-h_1-h_i, kh_1-1])$.
The inductive hypothesis guarantees that such a choice of the $c_j$ is possible, since we are only requiring convergence
of finitely many functions.

From Lemma \ref{pL} we know that the left side of $\Pi \cdot \eqref{ope3}$ contains powers of $z_1$ only
up to $z_1^{(k-2)h_1}$.  Now we want to isolate all negative powers of $z_1$ on the right side without
disturbing this condition.  We therefore rewrite $\eqref{ope3}$ in the following way:
\begin{equation} \label{big2}
\begin{split}
\IP{\prod_{a=1}^l V(\psi_a, z_a)}_{\eta,R} = & \sum_{i=2}^{l} \sum_{n = - h_1 - h_i}^{-1} \iota_{1,i} (z_1 - z_i)^{n} \IP{V(V_{-n-h_1}(\psi_1) \psi_i, z_i) \prod_{a \ne 1,i} V(\psi_a, z_a)}_{\eta,R} \\
- & \sum_{i=2}^{l-1} \sum_{n = - h_1 - h_i}^{-1} \iota_{i,1} (z_1 - z_i)^{n} \IP{V(V_{-n-h_1}(\psi_1) \psi_i, z_i) \prod_{a \ne 1,i} V(\psi_a, z_a)}_{\eta,R} \\
+ &\sum_{n = 0}^\infty (z_1 - z_l)^n \IP{V(V_{-n-h_1}(\psi_1) \psi_l, z_l) \prod_{a \ne 1,l} V(\psi_a, z_a)}_{\eta,R},
\end{split}
\end{equation}
which on choosing $c_1 = h_1$ and multiplying by $\Pi$ becomes
\begin{equation} \label{big1}
\begin{split}
\Pi \cdot \IP{\prod_{a=1}^l V(\psi_a, z_a)}_{\eta,R} = & \prod_{m=1}^k (z_1 - u_m)^{h_1} \sum_{i=2}^{l} \sum_{n = - h_1 - h_i}^{-1} \iota_{1,i} (z_1 - z_i)^{n} g_{i,n}(z_2, \dots, z_l) \\
-& \prod_{m=1}^k (z_1 - u_m)^{h_1} \sum_{i=2}^{l-1} \sum_{n = - h_1 - h_i}^{-1} \iota_{i,1} (z_1 - z_i)^{n} g_{i,n}(z_2, \dots, z_l) \\
+& \prod_{m=1}^k (z_1 - u_m)^{h_1} \sum_{n = 0}^\infty (z_1 - z_l)^n g_{l,n}(z_2, \dots, z_l).
\end{split}
\end{equation}
On the right side of \eqref{big1} all of the negative powers of $z_1$ have now been collected into the first term.
This term is
convergent in $R$ by our inductive hypothesis on the $g_{i,n}$.
So restrict attention
to the other two terms; write their sum $f(\zz)$,
\begin{equation} \label{deff}
\begin{split}
f(\zz) = -& \prod_{m=1}^k (z_1 - u_m)^{h_1} \sum_{i=2}^{l-1} \sum_{n = - h_1 - h_i}^{-1} \iota_{i,1} (z_1 - z_i)^{n} g_{i,n}(z_2, \dots, z_l) \\
+& \prod_{m=1}^k (z_1 - u_m)^{h_1} \sum_{n = 0}^\infty (z_1 - z_l)^n g_{l,n}(z_2, \dots, z_l),
\end{split}
\end{equation}
which is still a formal Laurent series.  

As remarked earlier, the left side of \eqref{big1}
only contains powers of $z_1$ up to $z_1^{(k-2)h_1}$; and it is clear that the first term on the right 
contains powers of $z_1$ only up to $z_1^{kh_1-1}$; so $f(\zz)$ is actually a polynomial in $z_1$,
of degree at most $kh_1 - 1$.  We can therefore expand $f(\zz)$ as
\begin{equation} \label{taylor}
f(\zz) = \sum_{s=0}^{kh_1-1} \frac{(z_1-z_l)^s}{s!} \left(\dwrt{z_1}\right)^s \Big\vert_{z_1 = z_l} f(\zz).
\end{equation}
To exploit \eqref{taylor} we must make the formal substitution $z_1 = z_l$ in each term of \eqref{deff}.  
To verify that this is well defined we need to check that, for
fixed $j_2, \dots, j_{l-1}$ and fixed $\alpha+\beta$, 
there are only finitely many terms $z_1^{\alpha} z_2^{j_2} \cdots z_{l-1}^{j_{l-1}} z_l^\beta$
appearing in each term of \eqref{deff}.  This in turn amounts to checking that the 
Laurent series $g_{i,n}$ has only a finite singularity at
 $z_l = 0$, with order bounded uniformly in $n$; this is automatic for $i \ne l$, and
for $i=l$ it is guaranteed by the
the fact that $z_l$ is the coordinate closest to the origin, by an argument similar
to that in the proof of Lemma \ref{opelem}.  So we can substitute \eqref{deff} into
\eqref{taylor}, obtaining finally a polynomial in $z_1$ whose coefficients are convergent power series in $R$
(by our inductive hypothesis on the relevant $g_{i,n}$.)  
So $f(\zz)$ is convergent in $R$.  Since we already dealt with the first term in \eqref{big1}, this
proves that $\Pi \cdot \IP{\prod_{a=1}^l V(\psi_a, z_a)}_{\eta,R}$ converges to a holomorphic function in $R$.
Call this function $\Pi \cdot \IP{\prod_{a=1}^l V(\psi_a, z_a)}_{\eta}$.

By \eqref{big2} it is clear that the recursive formula 
\eqref{ope5} is satisfied.  On the other hand, \eqref{ope5} defines a meromorphic function on all of $\C^l$,
so we get the required analytic continuation of the power series $\Pi \cdot \IP{\prod_{a=1}^l V(\psi_a, z_a)}_{\eta,R}$
to all of $\C^l$.  It only remains
to check that the resulting function is actually independent of which region $R$ we started with.

First suppose $R'$ is obtained from $R$ by swapping $z_i$ with $z_j$, for some $i,j \ne l$.  Choose any $m \ne l$.  Using
$\Pi \cdot$ \eqref{ope} and bringing $z_m$ close to $z_l$ we get the ``simple OPE'' for the functions analytically continued from $R$:
\begin{equation}
\Pi \cdot \IP{\prod_{a=1}^l V(\psi_a, z_a)}_{\eta,R} = \Pi \cdot \sum_{n = - h_m - h_l}^\infty (z_m - z_l)^{n} \IP{V(V_{-n-h_m}(\psi_m) \psi_l, z_l) \prod_{a \ne m,l} V(\psi_a, z_a)}_\eta
\end{equation}
But note that we would get the same thing on the right side had we started with $R'$ instead of $R$.  The analytic continuations from $R$ and $R'$ therefore
agree when $z_m$ is close to $z_l$, hence everywhere.

On the other hand, suppose $R'$ is obtained from $R$ by swapping $z_l$ with $z_{l-1}$.  In this case we need to use \eqref{opeshort} and the case $m=l-1$ of \eqref{ope};
multiplying both equations by $\Pi$ we get two expressions for $\Pi \cdot \IP{\prod_{a=1}^l V(\psi_a, z_a)}_{\eta,R}$ when $z_l$ is near $z_{l-1}$, namely
\begin{gather}
\Pi \cdot \IP{\prod_{a=1}^l V(\psi_a, z_a)}_{\eta,R} = \Pi \cdot \sum_{n = - h_{l-1} - h_l}^\infty (z_{l-1} - z_l)^{n} \IP{V(V_{-n-h_{l-1}}(\psi_{l-1}) \psi_l, z_l) \prod_{a \ne l-1,l} V(\psi_a, z_a)}_\eta, \\
\Pi \cdot \IP{\prod_{a=1}^l V(\psi_a, z_a)}_{\eta,R} = \Pi \cdot \sum_{n = - h_{l-1} - h_l}^\infty (z_{l} - z_{l-1})^{n} \IP{V(V_{-n-h_l}(\psi_l) \psi_{l-1}, z_{l-1}) \prod_{a \ne l-1,l} V(\psi_a, z_a)}_\eta.
\end{gather}
Exchanging the label $z_{l-1}$ for $z_l$ in one of the two equations makes manifest that the functions 
continued from $R$ and $R'$ agree when $z_l$ is near $z_{l-1}$, hence everywhere.
This completes the proof since we can transform any $R$ 
to any $R'$ by successive swaps of the types we have considered. \done

Conbining the last two lemmas, now we can finally prove that the original power series are well behaved, thus
establishing the existence of the correlation functions:

\begin{thm} \label{bigThm} Suppose given $\eta: \F \to \C$ such that $O_\uu \subset \ker \eta$.  Then the power
series $\IP{\prod_{a=1}^l V(\psi_a, z_a)}_{\eta,R}$ defined by \eqref{defseries} each converge on some domain and can be analytically continued to a single
function $\IP{\prod_{a=1}^l V(\psi_a, z_a)}_{\eta}$ which is meromorphic on $\C^l$.  This function has poles only at $z_i = z_j$ or $z_i = u_j$,
and $\IP{\prod_{a=1}^l V(\psi_a, z_a)}_{\eta} dz_i^{\otimes h_i}$ is nonsingular at 
$z_i = \infty$.  Furthermore, for all $i,j \in [1,l]$ we have the operator product expansion
\begin{equation}
\IP{\prod_{a=1}^l V(\psi_a, z_a)}_{\eta} = \sum_{n= - h_i - h_j}^\infty (z_i - z_j)^n \IP{V(V_{-n-h_i}(\psi_i) \psi_j, z_j) \prod_{a \ne i,j} V(\psi_a, z_a)}_\eta.
\end{equation}
for $z_i$ sufficiently close to $z_j$.
\end{thm}
\oproof{} This all follows directly from Lemma \ref{modconv} except for the behavior at $\infty$, which is a consequence of Lemma \ref{pL}. \done

Finally we can establish a limited form of 
the ``representation property'' in the sense of \cite{GG} (see also \cite{Montague}):

\begin{thm} Suppose given $\eta: \F \to \C$ such that $O_\uu \subset \ker \eta$.  Let $\OO$ be an open disc $\{ |z| < R \} \subset \C$, with all $u_i \notin \OO$.
Then there exists a state $\Sigma \in \V^\OO$ which induces $\eta$ in the sense that, for $\chi \in \F$,
\begin{equation}
\eta(\chi) = \IP{\Sigma \chi}.
\end{equation}
\end{thm}
\oproof{}  The topological vector space $\V^\OO$ contains the Fock space at $\infty$
which we denote $\F_\infty$ (to distinguish it from $\F$ which is the Fock space at $0$); the idea of the proof is to build up $\Sigma$ as a limit
of states in $\F_\infty$, using the existence of correlation functions to establish convergence.

We have $\F_\infty \subset \F^*$ via the rule \cite{Greview}
\begin{equation}
\psi(\chi) = \lim_{z \to 0} \IP{V((-z)^{-2 L_0} e^{-z^{-1} L_1} \psi, z^{-1}) V(\chi, z)} \qquad \forall \ \psi \in \F_\infty, \chi \in \F
\end{equation}
(note that this indeed defines an injection --- the definition of $\F_\infty$ by factorization guarantees that if any $\psi \in \F_\infty$ annihilates
every $\chi \in \F$ then $\psi = 0$.)  Both $\F$ and $\F_\infty$ are graded by conformal weight.
Writing $\F^{(N)}$ for the space of states with weight $\le N$, and likewise $\F^{(N)}_\infty$, 
we have $\dim \F^{(N)} = \dim \F^{(N)}_\infty$ (by assumption as described
in Section \ref{notation}, both dimensions are finite.)  
Let $P_N: \F \to \F^{(N)}$ denote the projection.  Then we claim that its adjoint $P_N^*: \F^* \to \F^*$
actually maps $\F^* \to \F^{(N)}_\infty$.  To prove this claim, note that $\F^{(N)}_\infty$ is contained in $P_N^*(\F^*)$,
since for any $\psi \in \F^{(N)}_\infty$, we have $\psi(\chi) = 0$ when $h_\chi > N$.  On the other hand the two spaces have equal 
dimension, which proves the claim.

Now write $\Sigma_N = P_N^* \eta$.  By the above, the $\Sigma_N$ are actually elements of $\F_\infty$.  
We claim that they converge as $N \to \infty$ to some $\Sigma \in \V^\OO$.  To check this 
we need only verify that, for any $(\psi_1, \dots, \psi_l) \in V^{\otimes l}$ and $\zz$ in some compact $K \subset \OO^k$ bounded away from the diagonals,
\begin{equation}
f_N(\zz) = \IP{\Sigma_N \prod_{a=1}^l V(\psi_a, z_a)} \to f(\zz) = \IP{\prod_{a=1}^l V(\psi_a, z_a)}_\eta
\end{equation}
uniformly on $K$.
So for fixed $\zz$, consider the function $\lambda \mapsto f(\lambda \zz)$ 
where $\lambda$ ranges over $\C^\times$.
This function is holomorphic on some disc containing $0 < \abs{\lambda} \le 1$; and now we claim that 
the $f_N(\zz)$ are nothing but the partial sums in its Laurent expansion, evaluated at $\lambda=1$.
To prove this claim we note that $f(\zz)$ and $f_N(\zz)$ satisfy the same operator product expansion identities;
using these identities we can reduce to the case $l=1$, in which case the claim is straightforward.
Hence the $f_N(\zz)$ converge to $f(\zz)$, and since the $\zz$-dependence in $f(\zz)$ is of a particularly simple sort
($f$ is a rational function of $\zz$, with poles only on the diagonals or at the $u_i$) it is clear that the
convergence of this Laurent expansion is uniform in $\zz$ in the required sense. \done

\section{Functional dependence} \label{funcdep}

In the last section we 
checked that a linear functional $\eta: \F \to \C$ vanishing on $O_\uu$ is sufficient to determine a set of
correlation functions involving $k$ highest weight states fixed at $\uu$.  Next we want to show that these correlation
functions can in fact be extended to general $\uu$, in a manner consistent with the ``Knizhnik-Zamolodchikov'' differential
equations imposed by the rule $L_{-1} \mapsto \partial$ \cite{KZ}.  We will find that this can indeed be done, provided
that we impose a finiteness condition which in some sense expresses the existence of a null-vector.

By Theorem \ref{bigThm}, we know that to determine correlation functions at each point, 
it is sufficient to give linear functionals $\eta(\uu): \F \to \C$ such that each $\eta(\uu)$ annihilates
$O_\uu$.  We want to arrange that the correlation functions $\IP{}_{\eta(\uu)}$ corresponding to $\eta(\uu)$ satisfy the
appropriate KZ-type equations:  explicitly, what we require is that
\begin{equation}
\oint_{u_i} \IP{L(z) \chi}_{\eta(\uu)} dz = \dwrt{u_i} \IP{\chi}_{\eta(\uu)}.
\end{equation}
This differential equation implies a differential equation for the functionals $\eta(\uu)$, which
we will construct below; the remainder of this section is essentially devoted to checking that this
equation (which we view as a kind of ``parallel transport'' problem for the $\eta(\uu)$) admits solutions.

First we introduce a bit of notation.
Let $X$ denote the open set $\{ \uu = (u_1, \dots, u_k) \in \PP^k : u_i \ne u_j \ \forall i \ne j, u_i \ne 0 \ \forall i \}$.
Let $B$ denote the trivial vector bundle $X \times \F$ over $X$, and let $\Gamma(U,B)$ denote the space
of holomorphic sections of $B$ over any $U \subset X$.  (Since $B$ is an infinite-dimensional vector bundle
we should say what we mean by a ``holomorphic section:''  to be exact, we mean a holomorphic section of
some finite-dimensional subbundle.)  Then for each $\uu \in X$, let $O_\uu \subset
B_\uu \iso \F$ be the subspace we defined in \eqref{defO}.  Then let $O$ denote the sheaf
of holomorphic sections of the collection of spaces $O_\uu$ --- in other words, we define
\begin{equation}
\Gamma(U, O) = \left\{ s \in \Gamma(U, B) \ \Big\vert \ s(\uu) \in O_\uu \ \ \forall \uu \in U \right\}.
\end{equation}
It is not clear \ti{a priori}
that $O$ is a vector bundle (for example, different
$O_\uu$ could have different codimensions in $\F$).

What is the relation between the different spaces $O_\uu$?  Let us work informally for
a moment to see what we should expect.  Suppose we consider
some $\chi \in \Gamma(U, B)$, and introduce the notation $W(\phi, u)$ for
an insertion of a primary field $\phi$ (corresponding to some highest weight representation
of the theory) at the point $u \in \PP$.
Then, \ti{once we have defined the correlation functions for general $\uu \in X$},
we would expect to have
\begin{equation} \label{exp}
\begin{split}
\dwrt{u_i} \IP{\prod_{a=1}^k W(\phi_a, u_a) \chi(\uu)} & = \IP{W(L_{-1} \phi_i, u_i) \prod_{a \ne i} W(\phi_a, u_a) \chi(\uu)} \\
                                                       & + \IP{\prod_{a=1}^l W(\phi_a, u_a) \dwrt{u_i} \chi(\uu)}.
\end{split}
\end{equation}
(There is a potential notational confusion here:  we emphasize that $\chi(\uu)$ refers to an element 
of $B_\uu \iso \F$, which lives at $0 \in \PP$, and not some kind of ``field at $\uu$.'')
In particular, suppose in fact that $\chi \in \Gamma(U, O)$.
Then by the definition of $O$ the left side of \eqref{exp} should vanish identically.
On the right side, by the usual trick of reversing the contour, we could think of $L_{-1}$ as acting on
$\chi(\uu)$ instead of on $\phi_i$.  To do this we need to get rid of the contributions from the poles
in $(z-u_a)$ for $a \ne i$.  We can do this using the highest weight condition, which
guarantees that these poles have order at most $2$.  

So, for $i \in [1,l]$ and $\uu \in X$, define an 
operator $L^i(\uu): \F \to \F$ as follows:  
\begin{equation} \label{defL}
L^i(\uu) = -\oint_0 dz L(z) f^i_\uu(z)
\end{equation}
where $f^i_\uu(z) dz$ is holomorphic on $\PP - \{0\}$, with 
\begin{equation} \label{fcond}
\nu_{u_j} ((f^i_\uu - \delta^{ij}) dz) \ge 2
\end{equation}
(so $f^i_\uu - 1$ has a zero of order $2$ at $u_i$, and $f^i_\uu$ has a zero of order $2$ at all $u_j$ with $j \ne i$.) 
The definition \eqref{defL} of $L^i(\uu)$ depends on which function $f^i_\uu$
we pick, but from the definition \eqref{defO} of $O_\uu$ we see at once that different choices only differ by maps $\F \to O_\uu$.
Now \eqref{exp} says that
\begin{equation}
0 = \IP{\prod_{a=1}^k W(\phi_a, u_a) \left( L^i(\uu) + \dwrt{u_i} \right) \chi(\uu)},
\end{equation}
in other words, $\left( L^i(\uu) + \dwrt{u_i} \right) \chi(\uu)$ is orthogonal to all highest weight states.
Writing 
\begin{equation} \label{defcon}
D^i(\uu) = L^i(\uu) + \dwrt{u_i},
\end{equation}
the above considerations lead us to expect:

\begin{lem} \label{dpres} For open sets $U \subset X$,
\begin{enumerate}
\item The operator $D^i: \Gamma(U,B) \to \Gamma(U,B)$ maps $\Gamma(U,O) \to \Gamma(U,O)$.
\item The operator $[D^i, D^j]$ maps $\Gamma(U,B) \to \Gamma(U,O)$.
\end{enumerate}
\end{lem}
\oproof{} First we fix a particular choice of $f^i_\uu(z)$ which will make the calculations easier.
Namely, we let $f^i_\uu(z)$ be of the form
\begin{equation}
f^i_\uu(z) = \frac{1}{z^{3k+1}} \prod_{j \ne i} (z-u_j)^3 (Az^2 + Bz + C)
\end{equation}
where $A$, $B$, $C$ are fixed by requiring that $f^i_\uu(z) - 1$ have a zero of order $3$ at $z = u_i$.
The point of this choice is that it makes $f^i_\uu$ holomorphic in $\uu$ so long as $\uu$ stays in $X$,
and satisfies 
\begin{equation} \label{cond3}
\nu_{u_j} \left((f^i_\uu - \delta^{ij}) dz \right) \ge 3.
\end{equation}

\medskip

\oproof{ of 1} First note we are indeed free to choose $f^i_\uu(z)$ as above, since different choices of $f^i_\uu(z)$
satisfying \eqref{fcond} change $D^i$ only by maps into $O_\uu$.
Now $\Gamma(U,O)$ is spanned (over the holomorphic functions on $U$) by sections of the form
\begin{equation} \label{sectdef}
s(\uu) = - \ci dw g(\uu,w) V(\psi,w) \chi,
\end{equation}
where $\chi \in \F$, $\psi \in V$ and $g(\uu, w)$ is holomorphic in $\uu$
(as given e.g. by \eqref{fchoice}).  (Strictly speaking, it is not completely obvious that sections of this form are enough ---
pathologies are excluded by choosing a maximal set
of sections \eqref{sectdef} which are linearly independent at a single point $\uu$, then noting that the subset of $U$ 
on which they become degenerate is nowhere dense.)
So it is sufficient to check that $D^i$ maps the section \eqref{sectdef} into a section of $O$.  We therefore compute 
\begin{equation} \label{p1}
\begin{split}
D^is(\uu) &= - \left(\dwrt{u_i} - \oint_{\abs{z} > \abs{w}} dz L(z) f^i(\uu, \zz) \right) \ci dw g(\uu, w) V(\psi, w) \chi \\
          &= - \ci dw \left(\dwrt{u_i} g(\uu,w) \right) V(\psi,w) \chi + \ci \ci_{\abs{z} > \abs{w}} dw dz f^i(\uu, z) g(\uu, w) L(z) V(\psi, w) \chi
\end{split}
\end{equation}
By the usual contour manipulation argument, the last term in \eqref{p1} can be rewritten as
\begin{equation} \label{p2} 
\begin{split}
& \ci_{\abs{w} > \abs{z}} dw g(\uu,w) V(\psi,w) \left(\ci dz f^i(\uu,z) L(z) \chi\right) \\
& + \ci_0 dw g(\uu,w) \ci_w dz f^i(\uu,z) \left(\frac{V(L_0 \psi, w)}{(z-w)^2} + \frac{V(L_{-1}\psi, w)}{z-w} + O((z-w)^0) \right) \chi,
\end{split}
\end{equation}
where in the second term we have used the OPE between $L(z)$ and $V(\psi, w)$.  Now the first term in \eqref{p2} is manifestly in $O_\uu(\F)$.  
Evaluating the integral of $z$ around $w$ in the second term we obtain
\begin{equation} \label{p3}
\ci_0 dw g(\uu,w) \left(f^i(\uu,w) V(L_{-1}\psi, w) + \dwrt{w} f^i(\uu,w) V(L_0 \psi, w) \right) \chi.
\end{equation}
Now the second term of \eqref{p3} is in $O_\uu(\F)$, but the first is not, because $L_{-1}\psi$ has weight $h_\psi+1$ and $gf^i(\uu,w)$
has a zero of order only $h_\psi$ at $w = u_i$.  
Combining it with the first term in \eqref{p1}, we see that what remains to be checked is that
\begin{equation} \label{p4}
\ci_0 dw  g(\uu,w) f^i(\uu,w) V(L_{-1} \psi, w) \chi - \left( \dwrt{u_i} g(\uu,w) \right) V(\psi,w) \chi
\end{equation}
belongs to $O_\uu(\F)$.  Using the fact that $V(L_{-1} \psi, w) = \dwrt{w} V(\psi, w)$ and integrating by parts, this boils
down to the assertion that
\begin{equation}
\dwrt{u_i} g(\uu,w) + \dwrt{w} \left(g(\uu,w) f^i(\uu,w) \right)
\end{equation}
has a zero of order at least 
$h_\psi$ at each $w = u_j$.  For $j \ne i$ this is clear since each term separately has such a zero.
At $w = u_i$ we use the fact that $f^i(\uu, w) = 1$ to second order in $w$. \done

\oproof{ of 2} Using the result of part 1, we see that we are again free
to use our convenient choice of $f^i_\uu(w)$.  With this choice we will show
that the operators $[L^j, \dwrt{u_i}]$ and $[L^i, L^j]$ \ti{separately} map $\Gamma(B,U) \to \Gamma(O,U)$ (for
other choices this would not be the case.)

So take any $\chi(\uu) \in \Gamma(B,U)$.  We have
\begin{equation}
\begin{split}
\left[L^j, \dwrt{u_i}\right] \chi(\uu) &= \dwrt{u_i} \ci dz L(z) f^i(\uu,z) \chi(\uu) - \ci dz L(z) f^i(\uu, z) \dwrt{u_i} \chi(\uu) \\
                            &= \ci dz L(z) \left(\dwrt{u_i} f^i(\uu,z)\right) \chi(\uu) 
\end{split}
\end{equation}
which belongs to $O_\uu$ by our hypothesis \eqref{cond3} on $f^i$ (this is
where we are using the fact that the number $3$ appears there, instead of the
$2$ in \eqref{fcond}.)  Next, for any $\chi \in \F$, we have the purely algebraic fact
\begin{equation}
\begin{split}
[L^i, L^j] \chi &= \left(\ci_{\abs{z}>\abs{w}} - \ci_{\abs{w}>\abs{z}}\right) dz dw f^j(\uu,z) f^i(\uu,w) L(z) L(w) \chi \\
	      	&= \ci_0 dw f^i(\uu, w) \ci_w dz f^j(\uu,z) \left( \frac{c/2}{(z-w)^4} + \frac{2L(w)}{(z-w)^2} + \frac{\partial_w L(w)}{z-w} + O((z-w)^0) \right) \chi \\
		&= \ci_0 dw \left( \frac{c}{2} f^i(\uu, w) \partial_w^3 f^j(\uu, w) + f^i(\uu,w) \partial_w f^j(\uu, w) L(w) + (i \leftrightarrow j) \right) \chi 
\end{split}
\end{equation}
and the first term vanishes since $f^i$, $f^j$ are holomorphic on $\PP - \{ 0 \}$, 
while the last two terms belong to $O_\uu$.  This completes the proof. \done

The $D^i$ as defined by \eqref{defcon} are the components of a connection (covariant derivative) in $B$, which
according to Lemma \eqref{dpres} is well defined and flat modulo sections of $O$.
So define the quotient sheaf $A = B/O$ by $\Gamma(U,A) = \Gamma(U,B) / \Gamma(U,O)$.
The $D^i$ then induce a flat connection in $A$
in the obvious way; we will use the letter $D$ for this connection as well.

To exploit the existence of this connection we will need to be able to solve differential equations
in the spaces of interest; to guarantee this can always be done, we now impose a strong
finiteness condition on the conformal field theory.  Namely consider the space 
$O_\uu$ at the point $\uu = (\infty, \dots, \infty)$.  If we call this space $C_k$, then \eqref{defO} becomes
\begin{equation} \label{defC}
C_k = \Span\left\{ V_{-N-(k-1)h}(\psi) \chi \ \Big\vert \ \chi \in \F, \psi \in V^h, N \ge 1 \right\}.
\end{equation}
Note that unlike the generic spaces $O_\uu$, $C_k$ inherits the grading from $\F$, so we can write $C_k = \oplus_{h \ge 0} C_k^h$.

We digress briefly to discuss the space $C_k$.  In the case $k=2$ it was originally introduced by Zhu in \cite{Zhu}, who
proved that the characters of the chiral theory close under modular transformations, under the 
hypothesis that $\F / C_2$ is finite-dimensional.  Zhu conjectured that this hypothesis is equivalent to
rationality of the theory.  As far as the author is aware, this conjecture is still unproven.

For our purposes the important point is that 
$C_k$ gives a kind of uniform control over the fibres of $O$, as we see from the following (essentially contained
already in \cite{Zhu} for $k=2$):

\begin{lem} \label{basislem} Let $S_k$ be a graded subspace of $\F$ with $S_k + C_k = \F$.  Then $S_k + O_\uu = \F$ for any $\uu \in X$.
\end{lem}
\oproof{}
First note that the term $V_{-N-(k-1)h}(\psi) \chi$ appearing in the definition \eqref{defC} is precisely 
the term of highest conformal weight
in the element of $O_\uu$ obtained by substituting $N, \psi, \chi$ in \eqref{fchoice}, so that any
element of $C_k$ equals an element of $O_\uu$ plus ``lower-order corrections.''  Explicitly, for any $M$, $C_k^{M} \subset O_\uu + \F^{(M-1)}$ (where by $\F^{(M-1)}$ we mean $\oplus_{h=0}^{M-1} \F^M$.)

Now we prove by induction that $\F^{(M)} \subset S_k^{(M)} + O_\uu$.  For $M=-1$ there is nothing to prove.  So assume $\F^{(M-1)} \subset S_k^{(M-1)} + O_\uu$.
By assumption we have $\F^M = S_k^M + C_k^M$, so $\F^M \subset S_k^M + O_\uu + \F^{(M-1)} \subset S_k^M + S_k^{(M-1)} + O_\uu = S_k^{(M)} + O_\uu$ as desired. \done

Now we can formulate our key finiteness hypothesis (which is a kind of higher-dimensional analogue of Zhu's ``condition C,'' to which
it reduces in case $k=2$) and our main lemma:
\begin{lem} \label{api} Suppose $\F / C_k$ is finite-dimensional.  Fix $\vv \in X$ and a simply connected neighborhood $U$ of $\vv$.  
Then any $\chi \in A_\vv$ may be extended to
$\chit \in \Gamma(U,A)$ such that $D^i \chit = 0$ for all $i \in [1,k]$.
\end{lem}
\oproof{} By assumption, we can find a finite-dimensional graded $S$ with $S + C_k = \F$.  Let $\{s_1, \dots, s_d\}$ be a basis for $S$.
From Lemma \ref{basislem} we know that $S + O_\uu = \F$ for all $\uu \in X$.  Let $U$ be any simply connected neighborhood of $\vv$ in $X$.  We write
\begin{equation}
\chit(\uu) = \sum_{l=1}^d f_l(\uu) s_l \quad (\Mod O_\uu)
\end{equation}
where the $f_l$ are complex-valued functions on $U$, yet to be determined.  Then
\begin{equation}
D^i \chit(\uu) = \sum_{l=1}^d \frac{\partial f_l}{\partial u_i}(\uu) s_l + f_l(\uu) L^i(\uu) s_l \quad (\Mod O_\uu).
\end{equation}
Writing $L^i(\uu) s_l = \sum_{m = 1}^d C^{im}_l(\uu) s_m \ (\Mod O_\uu)$, to get $D^i \chit(\uu) = 0$ it is therefore
sufficient to demand that
\begin{equation} \label{eqd}
0 = \frac{\partial f_m}{\partial u_i} + \sum_{l=1}^d f_l C^{im}_l (\uu)
\end{equation}
for each $m$.  For each $i$, this is a regular matrix
differential equation for the functions $f_l$; furthermore, the fact that $[D^i, D^j] = 0 \ (\Mod O_\uu)$ is exactly the integrability condition for this
system of differential equations, so Frobenius's theorem \cite{Spivak} implies they have a common solution with
the specified initial condition. \done

Now all the work has been done and we can prove our main theorem, which is essentially a translation of the last result
to the dual sheaf $A^*$.
\begin{thm} \label{mainthm} Suppose $\F / C_k$ is finite-dimensional.  Fix $\vv \in X$ and a simply connected neighborhood $U$ of $\vv$.
Then any $\eta \in A_\vv^*$ may be uniquely extended to $\etat \in \Gamma(A^*, U)$ such that for any $\chi \in \F$,
\begin{equation} \label{pop}
\etat(\uu)(L^i(\uu) \chi) = \dwrt{u_i} \etat(\uu) (\chi).
\end{equation}
\end{thm}
\oproof{} We define $\etat(\uu)$ by the following rule:  given any $\chi \in A_\uu$, use Lemma \ref{api} to extend $\chi$ to a section $\chit$ of $A$
over $U$ and in particular over $\vv$.  Then set $\etat(\uu)(\chi) = \eta(\chit(\vv))$.  The point of this definition is that it
makes $\etat$ ``constant on horizontal sections:'' for any covariantly constant section $\chit$ of $A$, we have
\begin{equation} \label{last}
\etat(\uu) (\chit(\uu)) = const.
\end{equation}
Differentiating \eqref{last} we obtain
\begin{equation} \label{last2}
\left( \dwrt{u_i} \etat(\uu) \right) (\chit(\uu)) + \etat(\uu) \left( \dwrt{u_i} \chit(\uu) \right) = 0,
\end{equation}
and since $D^i \chit(\uu) = 0$, using the definition \eqref{defcon} of $D$ in \eqref{last2} we obtain \eqref{pop}. \done

In terms of the correlation functions induced from $\etat(\uu)$ by Theorem \ref{bigThm}, the result \eqref{pop} can be reexpressed as
\begin{equation}
\oint_{u_i} \IP{L(z) \chi}_{\etat(\uu)} dz = \dwrt{u_i} \IP{\chi}_{\etat(\uu)},
\end{equation}
using the definition \eqref{defL} of $L^i$ and the fact that the correlation
functions satisfy the highest weight condition.
This is the desired functional dependence of the correlation functions on $\uu$.  So when $\F/C_k$ is finite-dimensional,
Theorem \ref{mainthm} (together with Theorem \ref{bigThm}) 
gives a construction of a complete family of $k$-point functions on simply connected neighborhoods in $X$, 
starting from a single linear functional on one
space $A_\vv$.

We remark that there is another approach to this system of differential equations, which gives a slightly different
result.  Namely, one can use the fact that all correlation functions $\IP{\prod_{a=1}^k W(\phi_a, u_a)}$ can be computed
in terms of correlations between fields $\phi_a$ belonging to the ``special subspaces'' \cite{Nahm2} of the $k$ relevant 
representations.  When at least $k-3$ of the representations are quasirational, one then finds that
the relevant differential equations close on a finite-dimensional space.  By a calculation similar
to that done in the proof of Lemma \ref{dpres} one
can then show that the integrability conditions are always satisfied, so
that the differential equations admit a solution.
It would be interesting to find a more explicit connection between this approach
and that presented above.

\section{Discussion} \label{discussion}
In this paper we have presented a construction of chiral correlation functions on the sphere.  This
construction guarantees that the correlation functions are locally single-valued, but tells us nothing
about the monodromy when two fields are transported around one another.  In generic situations one expects that the
correlation function will change at most by a phase under this transformation, but there are examples known in which this
is not the case, such as the \ti{logarithmic} conformal field theories \cite{log}.  It would be interesting to find
a natural condition which implies that logarithms do not occur.  In the case of $2$-point functions with
logarithms one sees at once that the problem is failure of $L_0$ to act semisimply; more generally 
it has been suggested \cite{Gpriv} that semisimplicity of Zhu's algebra might be sufficient to exclude
logarithmic behavior for all correlation functions (finite-dimensionality of $\F / C_2$ is not sufficient, as one
sees \cite{Gpriv} from the example of \cite{Glog}.)

After this work was completed the author became aware that vector bundles of conformal blocks with connection 
determined by the stress-energy tensor, 
similar to the sheaf $A^*$ appearing in Section \ref{funcdep}, were introduced by Friedan and Shenker and
have been considered previously in e.g. \cite{FS}, \cite{Felder},
\cite{FS2}.  These constructions are formulated on moduli spaces which are compactified by including
configurations in which the marked points come together; the vector bundle of conformal blocks becomes
nontrivial, and the connection only projectively flat, when these extra configurations are included.
It would be interesting to understand more precisely the relation between the Friedan-Shenker vector bundles
and those introduced in this paper; in particular, if there were a canonical way to extend the vector bundle $\F / O_\uu$
to points in moduli space where the $u_i$ coincide, it might shed light on the conjecture of Zhu mentioned in Section \ref{funcdep},
as well as the question of the monodromy of the correlation functions.

It would also be interesting to understand in more detail the
relation between the present work and the tensor product theory
of \cite{HL}.

\section*{Acknowledgements} \label{ack}

This paper could not have been written were it not for my research supervisors Peter Goddard and Matthias Gaberdiel,
who originally suggested the problem and who helped me negotiate numerous obstacles.  In particular, I would like to thank
Matthias Gaberdiel for several important suggestions and for helpful feedback on various drafts of the paper.  I would
also like to thank Sakura Sch\"afer-Nameki for her comments on an early draft.

I gratefully acknowledge financial support from a British Marshall Scholarship.

\end{document}